\documentclass[10pt,twocolumn,letterpaper]{article}

\usepackage{cvpr}              % To produce the CAMERA-READY version
\usepackage[accsupp]{axessibility}
\definecolor{cvprblue}{rgb}{0.21,0.49,0.74}
\usepackage[pagebackref,breaklinks,colorlinks,allcolors=cvprblue]{hyperref}
\usepackage{booktabs}
\usepackage{booktabs}
\usepackage{multirow}
\usepackage{makecell}
\usepackage{xcolor}
\usepackage{rotating}
\usepackage{caption}
\usepackage{geometry}
\usepackage{tikz}

\usepackage{tablefootnote}
\usepackage[table]{xcolor}  % enables \rowcolor for tables
\usetikzlibrary{shapes.geometric, arrows.meta, positioning, fit, backgrounds}
\hypersetup{
    colorlinks=true,
    citecolor=green
}

\definecolor{ours}{HTML}{D6EAF8}   % light blue highlight for our rows
\def\paperID{*****} % *** Enter the Paper ID here
\def\confName{CVPR}
\def\confYear{2026}

\title{TRACE: Training-Free Partial Audio Deepfake Detection via Embedding Trajectory Analysis of Speech Foundation Models}
\author{Awais Khan\\
University of Michigan-Flint, MI, USA\\ ProbeTruth Inc. \\
{\tt\small mawais@umich.edu}
\and
Muhammad Umar Farooq\\
University of Michigan-Flint, MI, USA\\ ProbeTruth Inc. \\
{\tt\small mufarooq@umich.edu}
\and
Kutub Uddin\\
University of Michigan-Flint, MI, USA,\\ 
{\tt\small kutub@umich.edu}
\and
Khalid Malik\\
University of Michigan-Flint, MI, USA\\ ProbeTruth Inc. \\
{\tt\small drmalik@umich.edu}
}
\begin{document}
\maketitle
% \vspace{-7mm}
\begin{abstract}
% \vspace{-7mm}
Partial audio deepfakes, where synthesized segments are spliced into genuine recordings, are particularly deceptive because most of the audio remains authentic. Existing detectors are supervised: they require frame-level annotations, overfit to specific synthesis pipelines, and must be retrained as new generative models emerge. We argue that this supervision is unnecessary. We hypothesize that speech foundation models implicitly encode a forensic signal: genuine speech forms smooth, slowly varying embedding trajectories, while splice boundaries introduce abrupt disruptions in frame-level transitions. Building on this, we propose \textbf{TRACE} (\textbf{T}raining-free \textbf{R}epresentation-based \textbf{A}udio \textbf{C}ountermeasure via \textbf{E}mbedding dynamics), a training-free framework that detects partial audio deepfakes by analyzing the first-order dynamics of frozen speech foundation model representations without any training, labeled data, or architectural modification. We evaluate TRACE on four benchmarks that span two languages using six speech foundation models. In \textit{PartialSpoof}, TRACE achieves 8.08\% EER, competitive with fine-tuned supervised baselines. In \textit{LlamaPartialSpoof}, the most challenging benchmark featuring LLM-driven commercial synthesis, TRACE surpasses a supervised baseline outright (24.12\% vs.\ 24.49\% EER) without any target-domain data. These results show that temporal dynamics in speech foundation models provide an effective, generalizable signal for training-free audio forensics.
\end{abstract}
    
\vspace{-7mm}
\section{Introduction}
\vspace{-2mm}
The rapid proliferation of neural text-to-speech and voice conversion systems has made high-quality speech synthesis accessible to non-experts, with modern models capable of cloning a voice from just a few seconds of audio~\cite{casanova2024xtts, liu2026ctc}. This capability has already been exploited in the real world, where cloned voices have been used to bypass speaker verification systems, facilitate financial fraud, and spread targeted disinformation~\cite{alali2025partial,lowy_audio_deepfakes}. These incidents show the urgent need for methods that remain effective as generative models continue to improve.\\
Early research in audio deepfake detection focused on fully synthesized utterances~\cite{tak2021end,khan2023spotnet,khan2023battling, uddin2025sheild, uddin2025advbench}, where an entire recording is generated by a single synthesis system. In this setting, supervised models~\cite{li2025audio,yang2026forensic,li2025survey,khan2024frame, uddin2025adversarial} operating on spectrograms or self-supervised speech embeddings achieve equal error rates well below 1\% on standard benchmarks, demonstrating that global distributional shifts between genuine and fully synthesized speech are comparatively easy to exploit~\cite{yi2022add, tak2021end}. However, these systems assume a homogeneous generative process across the entire utterance and are not designed to localize short manipulated regions embedded within bona fide speech. Consequently, they are easily fooled by more realistic threat: partial deepfakes~\cite{he2025manipulated}.\\
In a partial deepfake~\cite{zhang2021lcnn}, an adversary splices synthesized or out-of-context segments into a genuine recording to subtly alter its meaning while preserving the speaker identity for most of the clip. User studies and forensic evaluations show that humans detect such manipulations barely above chance and that commercial speaker verification systems can be misled with success rates exceeding 95\%~\cite{alali2025partial}. To address this, recent benchmarks such as PartialSpoof~\cite{zhang2022partialspoof}, HAD~\cite{Had-dataset}, and ADD 2023 Track 2~\cite{yi2023add} frame partial deepfake detection as a joint detection and localization problem, requiring systems to identify not only whether an utterance is manipulated but also which time spans have been altered. This setting exposes the limitations of detectors tuned to global cues and demands methods that operate at fine temporal granularity.\\
Current partial deepfake detectors~\cite{zhang2025partialedit,farooq2025transferable,martin2023vicomtech} are dominated by supervised deep learning built on top of large self-supervised speech models. While these systems achieve strong in-domain performance, they share three fundamental limitations. First, they require large amounts of frame-level annotated data, which is expensive and time-consuming to produce. Second, they tend to overfit to specific synthesis pipelines, leading to poor generalization when new generative models or editing tools emerge. Third, they must be repeatedly retrained or adapted as the threat landscape evolves, making deployment costly and fragile~\cite{he2025manipulated, li2025survey}. These constraints motivate a fundamentally different approach: one that leverages the intrinsic properties of pretrained speech foundation models without any task-specific learning.\\
Our key observation is that self-supervised speech foundation models~\cite{chen2022wavlm,baevski2020wav2vec,radford2023whisper,hsu2021hubert}, though never trained for forgery detection, implicitly encode a latent forensic signal. In the latent space of these models, bona fide speech traces smooth, slowly-varying trajectories governed by the continuity of human articulation and the shared acoustic context of a single speaker and recording environment. In contrast, splice boundaries break this continuity abruptly: the encoder must suddenly represent a segment produced by a different generative process, introducing a measurable disruption in the frame-level embedding transition rate. Our hypothesis is that this disruption is detectable through the first-order dynamics of consecutive frame-embedding distances, that is, how rapidly the representation changes from one frame to the next, without any model training or labeled data. We further investigate second-order dynamics in our ablation study.

Building on this hypothesis, we propose \textbf{TRACE} (\textbf{T}raining-free \textbf{R}epresentation-based \textbf{A}udio \textbf{C}ountermeasure via \textbf{E}mbedding dynamics), a training-free framework for partial audio deepfake detection that operates directly on frozen speech foundation model representations. TRACE requires no gradient updates, no annotated data, and no architectural modification, and can be applied uniformly across different backbone models and datasets. To our knowledge, this is the first study to demonstrate that pretrained speech foundation models can serve as effective forensic tools for partial audio deepfake detection in a training-free paradigm. Our main contributions are as follows.
\begin{itemize}
\item We identify the frame-level embedding transition rate as a training-free forensic signal in frozen speech foundation models and show empirically that splice boundaries produce measurable trajectory disruptions in the latent space of these models across multiple encoders and languages.
\item We propose TRACE, a training-free framework for partial audio deepfake detection that operates entirely on frozen speech foundation model representations without any labeled data, gradient updates, or architectural modification.
\item Through comprehensive evaluation on four benchmarks that span two languages and six foundation models, we show that TRACE achieves competitive performance with supervised detectors and surpasses baseline on LlamaPartialSpoof without any target-domain data, demonstrating that frozen speech foundation models are viable and effective forensic tools for training-free audio deepfake detection.
\end{itemize}
\vspace{-2.2mm}
\section{Related Work}
\vspace{-2mm}
\subsection{Partial Audio Deepfake Detection}
Audio deepfake detection has evolved from targeting fully synthesized utterances to addressing the more challenging partial manipulation setting. Early supervised models achieve sub-1\% EER on fully synthesized benchmarks by exploiting global distributional shifts between genuine and synthesized speech~\cite{yi2022add, tak2021end}. These systems, however, assume a homogeneous generative process and cannot localize short manipulated regions within otherwise bona fide speech. To address this limitation, dedicated benchmarks frame partial deepfake detection as a joint detection and localization problem: PartialSpoof provides frame-level annotations for TTS/VC-spliced English utterances~\cite{zhang2022partialspoof}, HAD constructs misleading half-truth statements from genuine phrases~\cite{Had-dataset}, ADD 2023 Track 2 evaluates systems under multiple unseen spoofing pipelines~\cite{yi2023add}, and LlamaPartialSpoof introduces LLM-driven commercial synthesis as a cross-domain stress test~\cite{luong2025llamapartialspoof}.\\
Supervised partial deepfake detectors fall into three families. Frame-level authenticity methods train classifiers on speech foundation model representations to label each frame as bona fide or spoofed~\cite{zhang2022partialspoof, fu2025multi, liu2025nes2net}. Boundary-perception methods, in contrast, explicitly model splice transitions: BAM achieves localization down to 160~ms via boundary-aware attention~\cite{zhong2024bam}, while TDL and W-TDL use embedding similarity and temporal convolution for frame-wise scoring~\cite{xie2024tdl, dragar2024w}. Similarly, inconsistency and dynamics-based models treat partial deepfakes as internally incoherent sequences: TDAM encodes first-order temporal differences of speech foundation model representations with directional attention, achieving state-of-the-art results on PartialSpoof and HAD~\cite{li2025tdam}, while GNCL and SAIL further incorporate graph consistency and semantic-acoustic reasoning~\cite{ge2025gncl, cao2025sail}. Despite strong in-domain results, all of these methods require frame-level annotated data, overfit to specific synthesis pipelines, and must be retrained as new generative models 
emerge~\cite{he2025manipulated, li2025survey}.
\vspace{-2.9mm}
\subsection{Training-Free and Embedding-Based Approaches}
\vspace{-2mm}
A complementary line of work reduces reliance on labeled data by exploiting frozen speech foundation model representations. Simple classifiers on top of frozen wav2vec, HuBERT, or XLS-R embeddings already deliver strong detection, confirming that these models encode rich forensic information without task-specific fine-tuning~\cite{lv2022fake, yi2022add}. Following this direction, latent-statistics approaches compute frame-level information entropy over frozen embeddings as a generalized deepfake indicator, though they still require limited 
fine-tuning~\cite{anshul2026av}. Retrieval-based methods such as TADA further reduce supervision by comparing test utterances against galleries of known generators in embedding space, enabling out-of-domain detection without retraining~\cite{stan2025tada}. Knowledge-guided approaches such as NE-PADD couple speech foundation model representations with named-entity knowledge to detect semantic-acoustic mismatches, but introduce complex auxiliary modules and external knowledge bases~\cite{xian2025nepadd}.\\
Our work is most closely related to TDAM~\cite{li2025tdam}, which also exploit the temporal dynamics of speech 
foundation model representations. However, TDAM is fully supervised: it trains a directional attention module on frame-level annotations and optimize task-specific objectives. In contrast, TRACE is strictly training-free: we argue that the first-order dynamics of frozen representations, specifically how rapidly the embedding trajectory changes between consecutive frames, already encode a sufficient forensic signal without any learning. Unlike retrieval-based approaches that require galleries of known examples~\cite{liu2025zeroday, stan2025tada}, TRACE operates on a single utterance with no external reference. To our knowledge, TRACE is the first framework to demonstrate that the geometric properties of frozen speech foundation models alone are sufficient for competitive partial audio deepfake detection across 
languages, synthesis methods, and unseen generative models.
\vspace{-2mm}
\section{Proposed Method: TRACE}
\begin{figure}[t]
  \centering
  \includegraphics[width=1\linewidth]{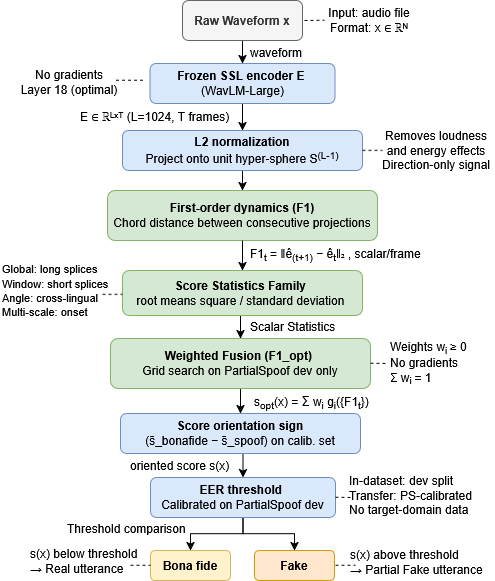}
  \caption{Overview of the TRACE pipeline. A raw waveform is passed through a frozen speech foundation model (WavLM-Large, layer~18). Frame embeddings are L2-normalized~\cite{van2017l2} onto the unit hypersphere, and the chord distance between consecutive projections forms the first-order dynamics sequence $\{\text{F1}_t\}$. Closed-form statistics are extracted and linearly fused into a scalar detection score, which is orientation-calibrated and threshold to produce the final bonafide or spoof decision. No model parameters are updated at any stage.}
  \label{fig:trace_pipeline}
  \vspace{-4mm}
\end{figure}

\vspace{-1mm}
TRACE detects partial audio deepfakes by analyzing the frame-level trajectory of frozen speech foundation model representations. Given a raw waveform, TRACE extracts embeddings from a frozen encoder~\cite{chen2022wavlm}, projects them onto the unit hypersphere, computes the chord distance between consecutive unit-sphere projections, and aggregates these into a scalar detection score via closed-form statistics. No model training, no labeled data, and no architectural modification is required at any stage. The flow diagram of the complete pipeline is presented in Figure~\ref{fig:trace_pipeline}.
\subsection{Embedding Extraction}
Let $\mathbf{x} \in \mathbb{R}^{N}$ be a raw waveform of $N$ samples. We pass it through a frozen pretrained speech foundation model $\mathcal{E}$ with a stride of 20\,ms (50\,Hz frame rate) to obtain a sequence of $L$-dimensional frame embeddings:%
\begin{equation}
    \mathbf{E} = \mathcal{E}(\mathbf{x}) = 
    \{\mathbf{e}_1, \mathbf{e}_2, \ldots, \mathbf{e}_T\} 
    \in \mathbb{R}^{L \times T},
\end{equation}
where $T$ is the number of frames and $L$ is the embedding dimension ($L{=}1024$ for WavLM~\cite{chen2022wavlm}). The encoder weights 
are entirely frozen: no fine-tuning, no gradient computation, and no architectural modification is applied at any stage.\\
\noindent\textbf{L2 normalisation.} Raw embedding magnitudes vary with loudness, recording level, and signal energy, all of which are independent of splice manipulation. To isolate the directional, phonological content of each frame from these confounds, we project each embedding onto the unit hypersphere:
\begin{equation}
    \hat{\mathbf{e}}_t = \frac{\mathbf{e}_t}{\|\mathbf{e}_t\|_2}, 
    \quad \hat{\mathbf{e}}_t \in \mathcal{S}^{L-1}.
\end{equation}
This normalization is critical: by operating on the unit hypersphere, all subsequent computations measure purely 
directional change, decoupled from recording-level magnitude variation and domain-specific energy differences.
\vspace{-1.3mm}
\subsection{First-Order Trajectory Dynamics}
\vspace{-1.3mm}
The central component of TRACE is the computation of the chord distance between consecutive unit-sphere projections. 
For each pair of adjacent frames, we compute:
\begin{equation}
    \text{F1}_t = \|\hat{\mathbf{e}}_{t+1} - 
    \hat{\mathbf{e}}_t\|_2, \quad t = 1, \ldots, T-1,
    \label{eq:f1}
\end{equation}
where $\|\hat{\mathbf{e}}_{t+1} - \hat{\mathbf{e}}_t\|_2$ is the chord distance in the ambient Euclidean space 
(not the geodesic $\arccos(\hat{\mathbf{e}}_t \cdot \hat{\mathbf{e}}_{t+1})$), yielding a simple closed-form 
scalar per frame. In bona fide speech, the resulting sequence $\{\text{F1}_t\}$ evolves smoothly, reflecting 
natural phonological transitions. At a splice boundary, the encoder representation shifts abruptly, producing a 
localized spike that would not arise in natural speech. By applying L2 normalisation~\cite{obi2023review} before differencing, TRACE 
measures purely directional change on the unit hypersphere, independent of magnitude, recording conditions, and 
speaker loudness. We show empirically in Section~\ref{sec:experiments} that this generalizes across 
languages and synthesis methods without retraining. We additionally examine second-order differences 
$\text{F2}_t = \text{F1}_{t+1} - \text{F1}_t$, but as shown in Section~\ref{sec:ablation}, F2 reduces to 
near-chance performance at the optimal encoder layer and is not used in the final system.
\subsection{Utterance-Level Score Statistics}
The frame-level sequence $\{\text{F1}_t\}_{t=1}^{T-1}$ is aggregated into a scalar utterance-level detection score 
via a summary statistic $g : \mathbb{R}^{T-1} \to \mathbb{R}$. We design four complementary families of statistics, each targeting a different structural property of partial deepfake attacks: global energy elevation for long spoof segments, peak localization for short segments, temporal onset detection for boundary transitions, and direction-invariant geometry for cross-lingual transfer.\\
\noindent\textbf{Base statistics} aggregate the F1 sequence globally and are effective when spoof segments are long 
relative to the utterance:
\begin{equation}
    s_{\text{rms}} = \sqrt{\frac{1}{T-1}\sum_t \text{F1}_t^2}, 
    \quad s_{\text{std}} = \sigma(\{\text{F1}_t\}), 
\end{equation}
    \begin{equation}
    \quad s_{\text{mean}} = \frac{1}{T-1}\sum_t \text{F1}_t.
\end{equation}

\noindent\textbf{Sliding-window maximum 
($\text{F1}_{\text{maxW}}$)} focuses on the most anomalous local window. When spoof segments are short or densely 
packed, global statistics dilute localized spikes. We therefore identify the single most suspicious window:
\begin{equation}
    s_{\text{maxW}} = \max_{w=1}^{T-W} 
    \sqrt{\frac{1}{W}\sum_{t=w}^{w+W-1} \text{F1}_t^2},
\end{equation}
where $W = 25$ frames (500\,ms). This is motivated by the observation that even a single splice boundary 
constitutes sufficient evidence of manipulation.\\
\noindent\textbf{Multi-scale derivatives ($\text{F1}_{\text{dt}k}$)} capture splice-induced onset patterns at different temporal resolutions by computing the RMS of $k$-step differences of the F1 sequence:
\begin{equation}
    s_{\text{dt}k} = \sqrt{\frac{1}{T-k}
    \sum_{t=1}^{T-k}(\text{F1}_{t+k} - \text{F1}_t)^2}, 
    \quad k \in \{1,2,3,4,5\}.
\end{equation}
\noindent\textbf{Directional angle statistics} provide a magnitude-independent complement by measuring the average 
angular deviation between consecutive embedding displacement vectors:
\begin{equation}
    s_{\text{angle}} = \frac{1}{T-2}\sum_{t=1}^{T-2}
    \arccos\!\left(\frac{\Delta\hat{\mathbf{e}}_t \cdot 
    \Delta\hat{\mathbf{e}}_{t+1}}{\|\Delta\hat{\mathbf{e}}_t\|_2 
    \|\Delta\hat{\mathbf{e}}_{t+1}\|_2}\right),
\end{equation}
where $\Delta\hat{\mathbf{e}}_t = \hat{\mathbf{e}}_{t+1} - \hat{\mathbf{e}}_t$. This statistic measures how much the 
direction of motion on the unit hypersphere changes between consecutive frame pairs. At a splice boundary, this direction shifts abruptly, producing a large angular deviation even when the global F1 magnitude is similar across domains, making it particularly valuable for cross-lingual generalization, as demonstrated in Section~\ref{sec:experiments}.
\vspace{-1.1mm}
\subsection{Score Combination and Calibration}
\vspace{-1.1mm}
\noindent\textbf{Score combination.} Different statistics capture complementary aspects of the trajectory disruption. 
We combine them via weighted linear fusion:
\begin{equation}
    s_{\text{opt}}(\mathbf{x}) = \sum_{i=1}^{K} w_i\, 
    g_i(\{\text{F1}_t\}), \quad w_i \geq 0, 
    \quad \sum_i w_i = 1,
\end{equation}
where weights $\{w_i\}$ are determined by exhaustive grid search over the PartialSpoof dev split with weight 
increments of 0.1, using EER as the sole selection criterion. No model parameters are updated and no gradients are computed. To guard against overfitting to the calibration set, we report both in-dataset EER using the calibrated combination and Transfer EER applying the PartialSpoof-calibrated combination directly to all other datasets without any adaptation, providing a strict measure of generalization. The optimal combination per dataset, denoted $\text{F1}_{\text{opt}}$, is reported in Section~\ref{sec:experiments}.

\noindent\textbf{Score orientation.} Different statistics and encoders produce scores with different natural 
directions: for some configurations a higher score indicates a spoof, while for others a lower score does. Rather than assuming a fixed direction, TRACE determines the correct orientation automatically from the calibration 
set via a single $\mathcal{O}(n)$ sign operation:
\begin{equation}
    \text{orientation} = 
    \text{sign}(\bar{s}_{\text{bonafide}} - 
    \bar{s}_{\text{spoof}}),
\end{equation}
where $\bar{s}$ denotes the mean score per class. If the sign is negative, indicating that spoof utterances score 
higher, we negate $s(\mathbf{x})$ before thresholding. This requires only the class membership labels of the 
calibration utterances and no task-specific annotations.\\
\noindent\textbf{Threshold selection.} For in-dataset evaluation, the EER threshold is determined on the calibration split. For cross-dataset transfer, the threshold calibrated on PartialSpoof dev is applied directly to all target datasets without adaptation, providing a strict measure of domain transferability with no target-domain data.
\vspace{-1.1mm}
\subsection{Computational Complexity}
The computational overhead of TRACE beyond the encoder forward pass is negligible. The F1 computation 
(Eq.~\ref{eq:f1}) is $\mathcal{O}(T \cdot L)$ and all summary statistics are $\mathcal{O}(T)$, requiring no 
gradients, no backpropagation, and no GPU memory beyond encoder inference. This makes TRACE practical for 
real-time deployment on standard hardware.
\vspace{-1mm}
\section{Experiments and Results}
\vspace{-1mm}
\label{sec:experiments}
\subsection{Experimental Setup}

\begin{table}[!t]
\centering
\footnotesize
\setlength{\tabcolsep}{3pt}
\renewcommand{\arraystretch}{1.05}

\caption{Statistics of the four partial audio deepfake detection benchmarks used in this work. BF=Bonafide, PF=Partial fake and FF= Fully Fake.}

\label{tab:data}

\begin{tabular}{llrrrr}
\toprule
\multirow{2}{*}{\textbf{Dataset}} &
\multirow{2}{*}{\textbf{Split}} &
\multicolumn{3}{c}{\textbf{Utterances}} &
\multirow{2}{*}{\textbf{Total}} \\

\cmidrule(lr){3-5}
& & \textbf{BF} & \textbf{PF} & \textbf{FF} & \\
\midrule

%% PartialSpoof
\multirow{3}{*}{\shortstack[l]{PartialSpoof\\\cite{zhang2022partialspoof}}}
& Train & 2580 & 22800 & -- & 25380 \\
& Dev   & 2548 & 22296 & -- & 24844 \\
& Eval  & 7355 & 63882 & -- & 71237 \\
\cmidrule(l){2-6}
& \textit{Total} & \textit{12483} & \textit{108978} & \textit{--} & \textit{121461} \\

\midrule

%% HAD
\multirow{3}{*}{\shortstack[l]{HAD\\\cite{Had-dataset}}}
& Train & 26554 & 26539 & -- & 53093 \\
& Dev   & 8914  & 8910  & -- & 17824 \\
& Test  & -- & 9072 & -- & 9072 \\
\cmidrule(l){2-6}
& \textit{Total} & \textit{35468} & \textit{44521} & \textit{--} & \textit{79989} \\

\midrule

%% ADD2023
\multirow{3}{*}{\shortstack[l]{ADD23 Tr.2\\\cite{yi2023add}}}
& Train & 26554 & 26539 & -- & 53093 \\
& Dev   & 8914  & 8910  & -- & 17824 \\
& Test  & 20000 & 30000 & -- & 50000 \\
\cmidrule(l){2-6}
& \textit{Total} & \textit{55468} & \textit{56539} & \textit{--} & \textit{120917} \\

\midrule

%% LlamaPartialSpoof
\multirow{2}{*}{\shortstack[l]{LlamaPS\\\cite{luong2025llamapartialspoof}}}
& R01TTS.0.a & 10573 & 32194 & 33461 & 76228 \\
& R01TTS.0.b & -- & -- & 64388 & 64388 \\
\cmidrule(l){2-6}
& \textit{Total} & \textit{10573} & \textit{32194} & \textit{97849} & \textit{140616} \\
\bottomrule
\end{tabular}
\vspace{-5mm}
\end{table}
\noindent\textbf{Datasets.} We evaluate TRACE on four partial audio deepfake benchmarks covering two languages and multiple splicing strategies (Table~\ref{tab:data}). \textit{PartialSpoof}~\cite{zhang2022partialspoof} is the 
standard English benchmark for partial spoof detection, containing frame-level annotations for TTS/VC-spliced 
utterances. \textit{HalfTruth Audio Deepfake (HAD)}~\cite{Had-dataset} and \textit{ADD 2023 Track 2}~\cite{yi2023add} are Mandarin benchmarks derived from AISHELL-3 and share identical train/dev splits; for HAD, we report results on the dev split since test-set bona fide labels are withheld by the organisers. \textit{LlamaPartialSpoof}~\cite{luong2025llamapartialspoof} is a recent English benchmark constructed using LLM-driven commercial TTS systems (i.e., ElevenLabs) and is treated strictly as a cross-dataset benchmark following the authors' protocol.

\noindent\textbf{Encoders.} We employ six pretrained speech foundation models spanning three architectural families: masked prediction with denoising WavLM-Large (317M parameters, 24 layers) and WavLM-Base (95M, 12 layers)~\cite{chen2022wavlm}; offline cluster-based prediction HuBERT-Large (317M, 24 layers)~\cite{hsu2021hubert}; contrastive learning Wav2Vec 2.0-Base (95M, 12 layers) and Wav2Vec 2.0-XLSR (300M, 24 layers)~\cite{baevski2020wav2vec}; and weak-supervised multitask ASR Whisper-Base (74M, 6 encoder layers)~\cite{radford2023whisper}. All models are used with entirely frozen weights. Embeddings are extracted 
at 50\,Hz frame rate (20\,ms stride).

\noindent\textbf{System configurations.} We evaluate four configurations of TRACE, all training-free: S1 uses WavLM-Large layer 24 (last layer) with $s_{\text{std}}$; S2 uses layer 18 with $s_{\text{rms}}$; S3 uses layer 18 with $s_{\text{maxW}}$; and the proposed system (Prop.) uses layer 18 with the optimized combination $\text{F1}_{\text{opt}}$.

\noindent\textbf{Evaluation metrics.} We report Equal Error Rate (EER, \%, $\downarrow$) and Area Under the ROC Curve (AUC, $\uparrow$). For cross-dataset evaluation, we additionally report Transfer EER: the EER obtained by applying the PartialSpoof dev-calibrated threshold directly to the target test set, strictly measuring domain transfer without any target-domain data. Free EER, the EER at the optimal threshold on the target set, serves as an upper bound on in-dataset 
performance.

\begin{table}[!b]
\vspace{-5mm}
\centering
\footnotesize
\caption{\textbf{TRACE performance on PartialSpoof dataset.}}
\label{tab:ps}
\vspace{-3mm}
\setlength{\tabcolsep}{2pt}
\renewcommand{\arraystretch}{0.92}
\begin{tabular}{lcccc}
\toprule
\textbf{System} & \textbf{Train} & \textbf{Res.} & \textbf{EER$\downarrow$} & \textbf{AUC$\uparrow$} \\
\midrule
CQCC-LCNN~\cite{Had-dataset} & PS & 20  & 27.17 & -- \\
LCNN-BLSTM~\cite{zhang2021initial} & PS & 160 & 12.84 & -- \\
SELCNN-BLSTM~\cite{zhang2021multi} & PS & 160 & 16.60 & -- \\
w2v2-large-5gMLP~\cite{zhang2022partialspoof} & PS & 160 & \underline{9.24} & -- \\
\midrule
S1 (L24,last) & -- & 20 & 16.37 & 0.91 \\
S2 (L18,F1-rms) & -- & 20 & 11.08 & \underline{0.95} \\
S3 (L18,F1-maxW) & -- & 20 & 14.68 & 0.92 \\
\textbf{Prop.} (L18,F1-opt) & -- & 20 & \textbf{8.08} & \textbf{0.97} \\
\bottomrule
\end{tabular}
\vspace{-2mm}
\end{table}

% ── TABLE 2: HAD ─────────────────────────────────────────────────────────────
\begin{table}[!b]
\vspace{-1mm}
\centering
\footnotesize
\caption{\textbf{TRACE performance on HAD and ADD~2023.}}
\vspace{-3mm}
\label{tab:had&add}
\setlength{\tabcolsep}{2pt}
\renewcommand{\arraystretch}{0.92}
\begin{tabular}{lccc}
\toprule
\textbf{System} & \textbf{Train} & \textbf{EER$\downarrow$} & \textbf{AUC$\uparrow$} \\
\midrule
\multicolumn{4}{l}{\textit{HAD~\cite{Had-dataset}}} \\
S1 (L24,last) & -- & 31.40 & 0.758 \\
S2 (L18,F1-rms) & -- & 30.11 & 0.765 \\
S3 (L18,F1-maxW) & -- & \underline{25.62} & \underline{0.821} \\
\textbf{Prop.} (L18,F1-opt) & -- & \textbf{20.92} & \textbf{0.869} \\
\midrule
\multicolumn{4}{l}{\textit{ADD~2023~\cite{yi2023add}}} \\
S1 (L24,last) & -- & 48.23 & 0.531 \\
S2 (L18,F1-rms) & -- & \textbf{33.43} & \textbf{0.738} \\
S3 (L18,F1-maxW) & -- & 34.81 & \underline{0.737} \\
\textbf{Prop.} (L18,F1-opt) & -- & \textbf{33.43} & \textbf{0.738} \\
\bottomrule
\end{tabular}
\vspace{-4mm}
\end{table}

\begin{figure*}[htb]
\centering
\includegraphics[width=0.80\linewidth]{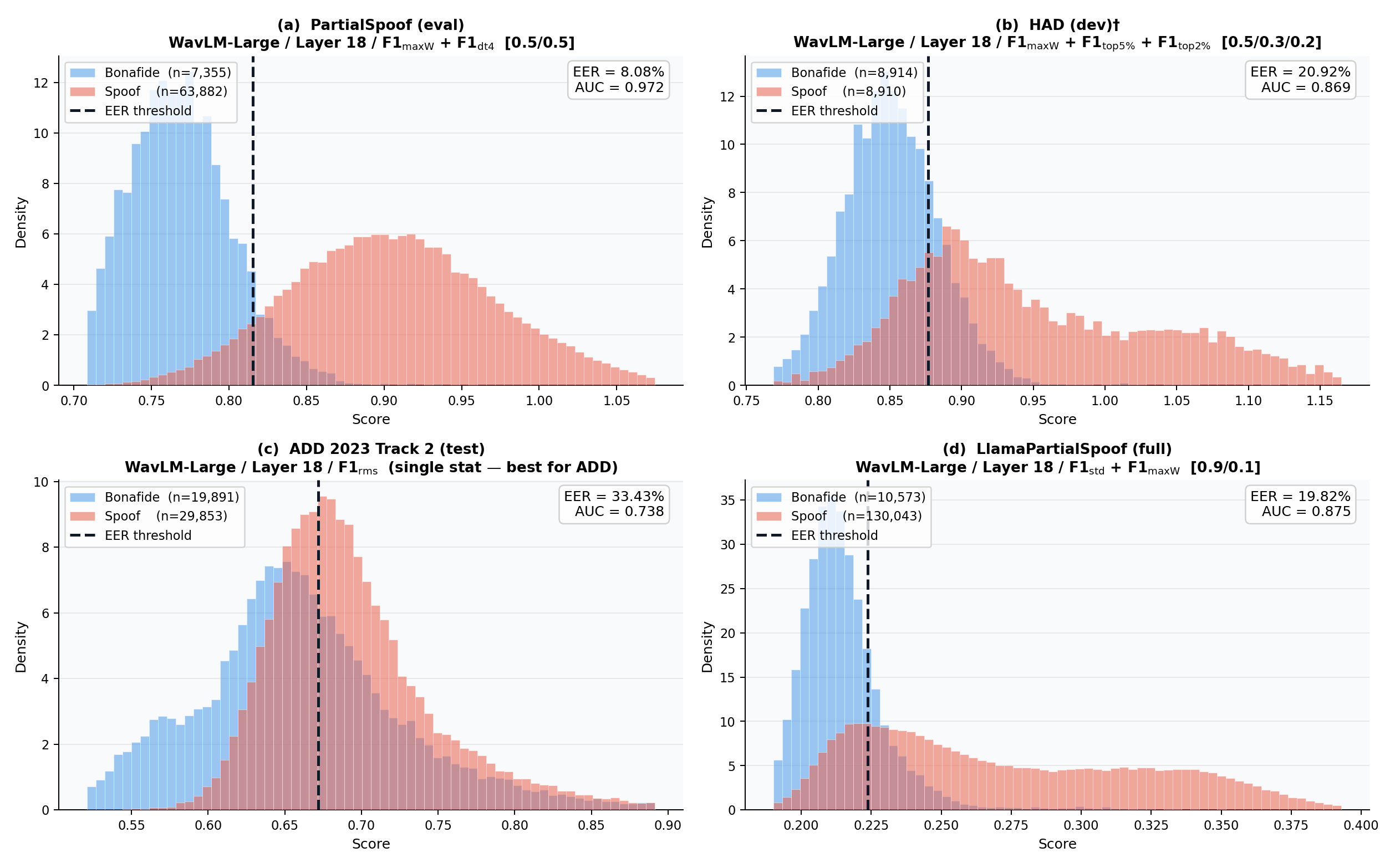}
\caption{
Score distributions of TRACE across four benchmarks: (a) PartialSpoof, (b) HAD, (c) ADD~2023, (d) LlamaPartialSpoof. The consistent directionality across datasets confirms language and synthesis-method independence of the TRACE.}
\label{fig:score_dist}
\vspace{-6mm}
\end{figure*}
\subsection{Main Results: PartialSpoof}
\label{sec:partialspoof}
Table~\ref{tab:ps} reports the performance of TRACE alongside supervised baselines on PartialSpoof. The proposed system achieves \textbf{8.08\% EER} and \textbf{AUC = 0.97}, competitive with supervised detectors that require frame-level annotated training data. Notably, TRACE outperforms the CQCC-LCNN (27.17\%)~\cite{Had-dataset}, LCNN-BLSTM (12.84\%)~\cite{zhang2021initial}, and LFCC-SELCNN-BLSTM (16.60\%)~\cite{zhang2021multi} baselines entirely without training, and approaches the fine-tuned SSL baseline w2v2-large-5gMLP (9.24\%)~\cite{zhang2022partialspoof}  while requiring no labeled data. The clean bimodal score separation visible in Figure~\ref{fig:score_dist}(a) confirms that the first-order embedding dynamics provide a reliable forensic signal on this benchmark.
\vspace{-2mm}
\subsection{Cross-Lingual Evaluation}
\label{sec:crosslingual}
To assess cross-lingual robustness, we evaluate TRACE on two Mandarin benchmarks: HAD~\cite{Had-dataset} and 
ADD 2023 Track 2~\cite{yi2023add}. Table~\ref{tab:had&add} summarizes the results.\\
On HAD, the proposed system achieves \textbf{20.92\% EER} (AUC = 0.869). The degradation relative to PartialSpoof reflects the denser splice structure of HAD, where multiple manipulated regions appear within a single utterance, causing global statistics to dilute localized anomalies. This is also visible in Figure~\ref{fig:score_dist}(b), where bona fide and spoof score distributions overlap more strongly than in the English benchmark. The sliding-window statistic $\text{F1}_{\text{maxW}}$ addresses this by focusing on the most anomalous 25-frame region, reducing EER 
from 30.11\% to 25.62\%, and further to 20.92\% when fused with complementary percentile statistics. \\
On ADD2023~\cite{yi2023add} (Table~\ref{tab:had&add}), spoof segments typically occupy a very small portion of the utterance. Under these conditions, aggressive localization risks capturing natural speech transitions, making global statistics more stable. Consequently, $s_{\text{rms}}$ achieves the best result of \textbf{33.43\% EER} (AUC = 0.738), consistent with the weaker score separation observed in Figure~\ref{fig:score_dist}(c). Across both datasets, intermediate SSL representations remain crucial: moving from the final transformer layer to layer 18 reduces ADD2023 EER from 48.23\% to 33.43\%. These results demonstrate that TRACE maintains cross-lingual effectiveness while adapting to different splice structures without any training on spoofed data.
\begin{table}[!b]
\centering
\footnotesize
\vspace{-4mm}
\caption{TRACE utterance-level EER (\%) on LlamaPartialSpoof by spoof type. Free EER uses subset-optimal thresholds; Transfer EER applies PartialSpoof-dev-calibrated threshold. Fully-fake utterances contain no splice and are inherently harder to detect.}
\vspace{-2mm}
\label{tab:llama_breakdown_short}
\setlength{\tabcolsep}{1.8pt}
\renewcommand{\arraystretch}{1.0}
\begin{tabular}{lrrr}
\toprule
\textbf{Spoof subset / System} & \textbf{Free EER$\downarrow$} & \textbf{Transfer EER$\downarrow$} & \textbf{AUC$\uparrow$} \\
\midrule
\multicolumn{4}{l}{\textit{TRACE (Ours)}} \\[1pt]
Crossfade partial-fake & 16.12 & 16.12 & 0.915 \\
Cut/paste + overlap-add & 13.04 & 12.60 & 0.940 \\
Fully-fake (TTS only) & 45.45 & 46.06 & 0.572 \\
\midrule
\multicolumn{4}{l}{\textit{Comparative SOTA (All spoof combined)}} \\[1pt]
w2v2-Large-MultiRes\cite{zhang2022partialspoof} & 47.49 & -- & -- \\
WavLM-BAM\cite{zhong2024bam} & 42.58 & -- & 0.534 \\
w2v2-XLsr-SAL\cite{mao2026localizing} & 35.52 & -- & 0.553 \\
WavLM-SAL\cite{mao2026localizing} & 36.60 & -- & 0.561 \\
PartialSpoof-trained baseline\cite{luong2025llamapartialspoof} & 24.49 & -- & -- \\
\textbf{TRACE (Ours, Full Corpus)} & \textbf{24.12} & \textbf{22.08} & \textbf{0.839} \\
\bottomrule
\end{tabular}
\vspace{-6mm}
\end{table}
\begin{figure*}[t]
  \centering
  \includegraphics[width=0.85\linewidth]{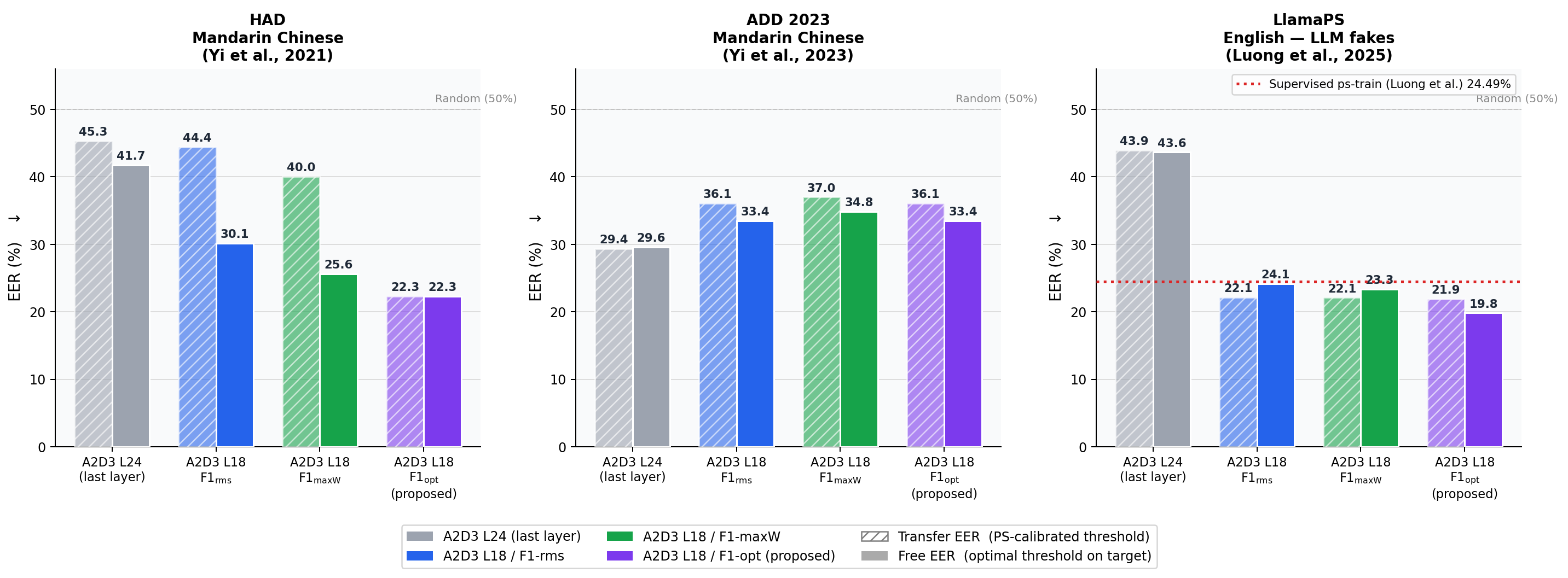}
  \caption{%
    Cross-dataset generalization of TRACE: Transfer EER (hatched) vs Free EER (solid) across three out-of-domain test sets. LlamaPartialSpoof includes the supervised ps-train baseline (red dashed) for reference.}
  \label{fig:cross_dataset}
  \vspace{-2mm}
\end{figure*}
\subsection{Cross-Corpus Generalization}
\label{sec:crosscorpus}
We evaluate cross-corpus generalization by calibrating all systems exclusively on PartialSpoof~\cite{zhang2022partialspoof} dev subset and apply the calibrated threshold directly to three out-of-domain test sets. Table~\ref{tab:llama_breakdown_short} and Figure~\ref{fig:cross_dataset} summarize the results.

\noindent\textbf{LlamaPartialSpoof.} On crossfade partial-fake utterances, TRACE achieves \textbf{16.12\% EER}, and on cut/paste and overlap-add utterances \textbf{13.04\% EER}. Fully-fake TTS utterances, which contain no splice boundary, yield higher EER (${\sim}$45\%), reflecting an expected limitation of boundary-focused statistics. Aggregating across the  full corpus, TRACE achieves \textbf{24.12\% Free EER} and \textbf{22.08\% Transfer EER}, surpassing the supervised PartialSpoof-trained baseline (24.49\%) without observing any LlamaPS~\cite{luong2025llamapartialspoof} data. This confirms that frozen speech foundation model embeddings capture robust partial-spoof cues that generalize to unseen 
LLM-driven forgeries.

\noindent\textbf{HAD and ADD 2023 transfer.} Applying the PartialSpoof-calibrated threshold directly to HAD shows a substantial Transfer vs Free EER gap (44.4\% vs 30.1\%), indicating score-scale mismatch due to cross-lingual domain shift. However, using the proposed $\text{F1}_{\text{opt}}$ combination, which mixes magnitude and direction-based features ($\text{F1}_{\text{maxW}}$ + $s_{\text{angle}}$, weights [0.7, 0.3]), reduces HAD Transfer EER to \textbf{22.3\%}, a 22\,pp improvement. The convergence of Transfer and Free EER bars in Figure~\ref{fig:cross_dataset} confirms that 
direction-invariant features bridge cross-domain gaps effectively. ADD 2023, which shares training material with HAD, aligns better with Transfer 36.1\% vs Free 
33.4\%.

\noindent\textbf{Language and domain independence.} 
Calibration on English speech and evaluation on Mandarin and LLM-synthesized English confirm that temporal continuity cues are language-agnostic: splice boundaries induce embedding discontinuities regardless of phonology, and the signal generalizes to unseen synthesis methods without retraining.

\subsection{Ablation Study}
\label{sec:ablation}
\begin{figure*}[t]
  \centering
  \includegraphics[width=0.80\linewidth]{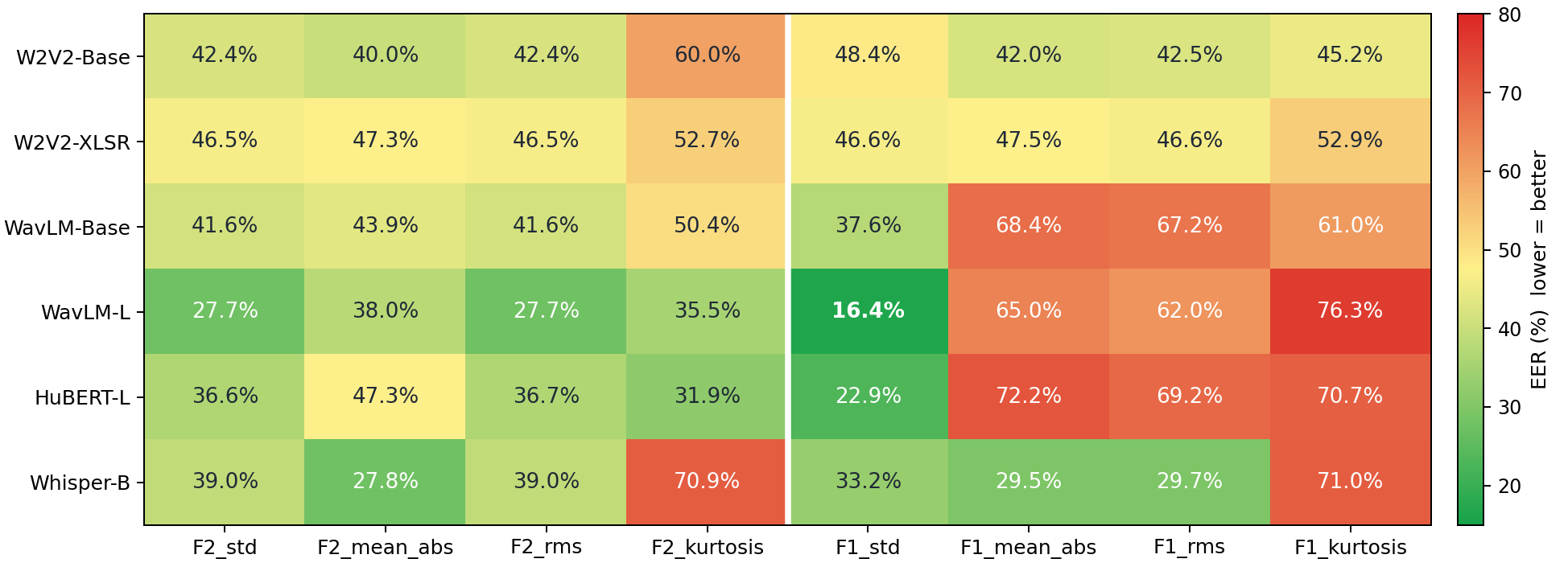}
  \caption{%
    \textbf{Encoder $\times$ statistic EER heatmap on PartialSpoof.} F1 consistently outperforms F2 across all encoders. WavLM-Large + F1-std achieves the best EER (16.4\%). Kurtosis-based features are unstable due to sensitivity to outliers.
  }
  \label{fig:eer_heatmap}
  \vspace{-6mm}
\end{figure*}

\vspace{-1.3mm}
\noindent\textbf{Encoder and statistic selection.} 
Figure~\ref{fig:eer_heatmap} shows EER across six encoders and eight base statistics on PartialSpoof~\cite{zhang2022partialspoof} dataset. WavLM-Large consistently achieves the lowest EER, reflecting superior representation quality from its masked prediction with denoising pretraining objective~\cite{chen2022wavlm}. First-order dynamics consistently outperform second-order dynamics across all encoders, confirming that the frame-level embedding 
transition rate carries more discriminative information than its rate of change. Contrastive models (Wav2Vec2.0~\cite{baevski2020wav2vec}) underperform mask-prediction models (WavLM~\cite{chen2022wavlm}, HuBERT~\cite{hsu2021hubert}), highlighting the importance of pretraining objectives that preserve temporal structure.

\begin{figure}[t]
  \centering
  \includegraphics[width=1.0\linewidth]{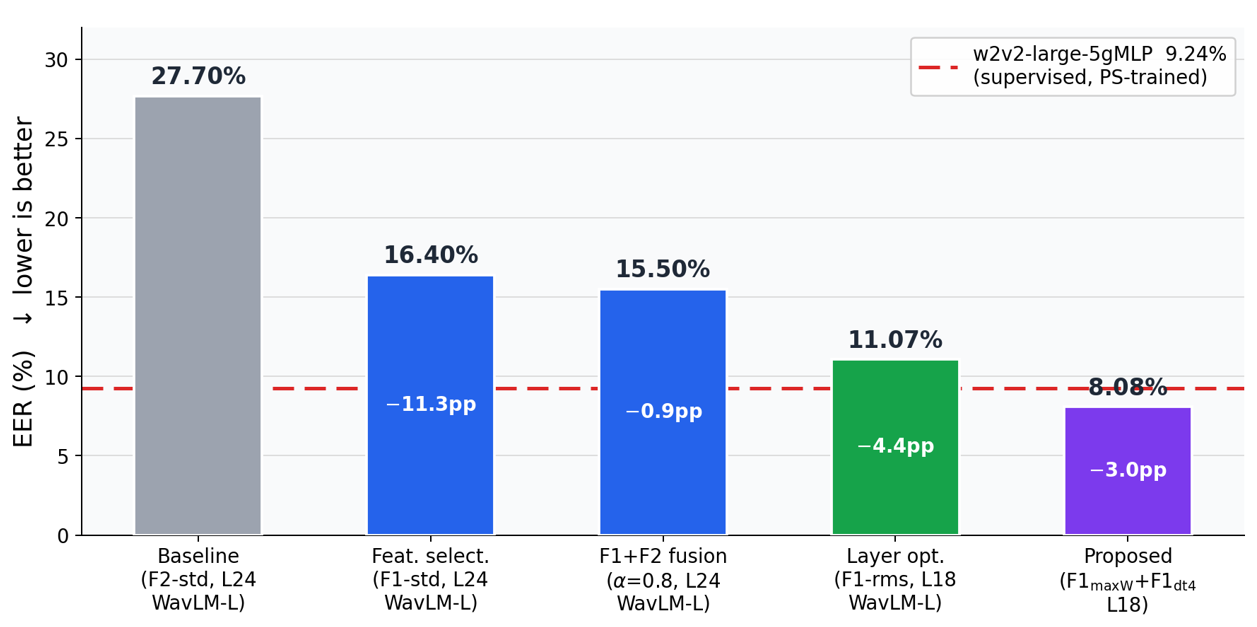} \vspace{-6mm}
  \caption{ Progressive improvement of TRACE on the PartialSpoof dataset. Horizontal dashed lines denote supervised baselines reported in \cite{zhang2022partialspoof}.}
  \label{fig:progressive}
  \vspace{-3.5mm}
\end{figure}

\noindent\textbf{Progressive design refinement.} 
Starting from the simplest baseline F2-std on WavLM-Large (layer 24) yields 27.7\% EER, already demonstrating that temporal dynamics in frozen speech foundation model representations encode partial-spoof cues. Switching to first-order dynamics (F1-std) reduces EER to 16.4\%. A layer ablation reveals that intermediate representations are more informative 
than the final transformer layer; using layer 18 with $s_{\text{rms}}$ further reduces EER to 11.1\%. Finally, the optimized feature combination ($\text{F1}_{\text{maxW}}$ + $\text{F1}_{\text{dt4}}$, equal weights) achieves \textbf{8.08\% EER}, yielding a 71\% relative error reduction from the initial baseline. This progression is illustrated in Figure~\ref{fig:progressive}.
%% Figure: Progressive improvement

\noindent\textbf{Feature statistics.} Among 43 evaluated statistics (details provided in supplementary materials), $s_{\text{rms}}$ (11.07\%), $s_{\text{mean\text{-}abs}}$ (10.84\%), and $\text{F1}_{\text{dt4\text{-}rms}}$ (11.08\%) perform strongest on PartialSpoof. Sliding-window variants are slightly weaker in-dataset but significantly better on cross-domain benchmarks. Directional angle features contribute little standalone but improve cross-domain generalization when combined with magnitude-based statistics. All F2 statistics remain near 50\% EER, confirming minimal utility for second-order dynamics at the optimal encoder layer.

\noindent\textbf{Feature combination.} Weighted combinations of complementary statistics consistently improve performance across all datasets, as summarized in Table~\ref{tab:combinations_short}. On PartialSpoof, $\text{F1}_{\text{maxW}}$ + $\text{F1}_{\text{dt4}}$ reduces EER from 11.07\% to 8.08\%. On HAD, combining three statistics reduces EER from 30.11\% to 20.92\%. On LlamaPS, $s_{\text{std}}$ + $\text{F1}_{\text{maxW}}$ lowers EER from 24.15\% to 19.82\% without any target-domain data, demonstrating that complementary embedding dynamics are both robust and generalizable across datasets and languages.
\begin{table}[t]
\centering
\footnotesize
\setlength{\tabcolsep}{2.2pt}
\renewcommand{\arraystretch}{1.0}
\caption{Best feature combinations per dataset (WavLM-Large, layer~18). $\Delta$ is improvement over F1-rms baseline.}
\begin{tabular}{lrrr}
\toprule
\textbf{Dataset} & \textbf{Best combination} & \textbf{EER\,\%} & $\Delta$ \\
\midrule
PartialSpoof~\cite{zhang2022partialspoof} & F1$_\text{maxW}$ + F1$_{\text{dt4}}$& 8.08 & $-$3.0\,pp \\
HAD~\cite{Had-dataset}          & F1$_\text{maxW}$ + F1$_\text{top5\%}$ + F1$_\text{top2\%}$  & 20.92 & $-$9.2\,pp \\
ADD~2023~\cite{yi2023add}     & F1$_\text{rms}$ (single) & 33.43 & -- \\
LlamaPS~\cite{luong2025llamapartialspoof}      & F1$_\text{std}$ + F1$_\text{maxW}$ & 19.82 & $-$4.3\,pp \\
\bottomrule
\end{tabular}
\vspace{-4mm}
\label{tab:combinations_short}
\end{table}
% \begin{figure}[!b]
%   \centering
%   \vspace{-4mm}
%   \includegraphics[width=\linewidth]{figures/fig_novel_stats.pdf}
%   \caption{%
%     \textbf{EER of 43 feature statistics on PartialSpoof (WavLM-Large, layer~18).} 
%     F1-rms, F1-mean-abs, and F1-dt4-rms perform best (11.07–11.17\%). Directional angle features alone are weak but aid cross-domain generalization when fused. F2 statistics score $\approx$50\% EER.
%   }
%   \label{fig:novel_stats}
%   \vspace{-4mm}
% \end{figure}
\vspace{-2mm}
\section{Conclusion}
\vspace{-2mm}
\label{sec:conclusion}
In this paper, we presented TRACE, a training-free framework for partial audio deepfake detection that operates directly on frozen speech foundation model representations. Our central hypothesis was that these models, though never trained for forgery detection, implicitly encode a latent forensic signal: bona fide speech traces smooth trajectories in their embedding space, whereas splice boundaries introduce abrupt disruptions measurable as the chord distance between consecutive unit-sphere projections. Our experiments across four benchmarks and six foundation models support this hypothesis. On PartialSpoof, TRACE achieves 8.08\% EER, competitive with supervised detectors that require frame-level annotations. More importantly, on LlamaPartialSpoof, TRACE surpasses a supervised baseline outright (24.12\% vs.\ 24.49\% EER) without any target-domain data, confirming that the forensic signal generalizes across languages, synthesis methods, and unseen generative models. \\
Our results further reveal three important insights. First, pretraining objectives that preserve temporal structure, such as masked prediction with denoising in WavLM-Large, are better suited for trajectory-based forensic analysis than contrastive objectives. Second, intermediate transformer layers are more informative than the final layer, suggesting that high-level semantic representations suppress the low-level acoustic discontinuities that TRACE exploits. Third, second-order dynamics collapse to near-chance performance at the optimal layer, confirming that first-order chord distances carry the dominant forensic information in frozen representations. We argue that analyzing the intrinsic behavioral signals of pretrained foundation models, rather than training on top of them, represents a promising and underexplored direction for scalable, data-independent audio forensics. As speech foundation models grow in scale, their latent representations are likely to encode increasingly rich forensic cues, making training-free approaches such as TRACE more effective over time. Two limitations remain: TRACE is designed around splice boundaries and performs poorly on fully synthesized utterances, and the statistic combination is calibrated on PartialSpoof dev, introducing dataset-specific dependency. Addressing both through universal, annotation-free statistic selection is a natural direction for future work.

\section{Acknowledgment}
This material is based upon work supported by the National Science Foundation (NSF) under Grant No.~2409577. Any opinions, findings, and conclusions or recommendations expressed in this material are those of the authors and do not necessarily reflect the views of the National Science Foundation.
{
    \small
    \bibliographystyle{ieeenat_fullname}
    \bibliography{main}

@String(ICASSP=	{ICASSP})

@String(IJCAI = {IJCAI})

@article{zhang2022partialspoof,
  author    = {Lin Zhang and Xin Wang and Erica Cooper and Nicholas Evans and Junichi Yamagishi},
  title     = {The PartialSpoof Database and Countermeasures for the Detection of Short Fake Speech Segments Embedded in an Utterance},
  journal   = {IEEE/ACM Transactions on Audio, Speech, and Language Processing},
  volume    = {31},
  pages     = {813--825},
  year      = {2022}
}

@article{liu2026ctc,
  title={CTC-TTS: LLM-based dual-streaming text-to-speech with CTC alignment},
  author={Liu, Hanwen and Yusuyin, Saierdaer and Huang, Hao and Ou, Zhijian},
  journal={arXiv preprint arXiv:2602.19574},
  year={2026}
}

@article{casanova2024xtts,
  title={Xtts: a massively multilingual zero-shot text-to-speech model},
  author={Casanova, Edresson and Davis, Kelly and G{\"o}lge, Eren and G{\"o}knar, G{\"o}rkem and Gulea, Iulian and Hart, Logan and Aljafari, Aya and Meyer, Joshua and Morais, Reuben and Olayemi, Samuel and others},
  journal={arXiv preprint arXiv:2406.04904},
  year={2024}
}

@inproceedings{khan2023spotnet,
  title={Spotnet: A spoofing-aware transformer network for effective synthetic speech detection},
  author={Khan, Awais and Malik, Khalid Mahmood},
  booktitle={Proceedings of the 2nd ACM International Workshop on Multimedia AI against Disinformation},
  pages={10--18},
  year={2023}
}

@inproceedings{khan2024frame,
  title={Frame-to-utterance convergence: A spectra-temporal approach for unified spoofing detection},
  author={Khan, Awais and Malik, Khalid Mahmood and Nawaz, Shah},
  booktitle={ICASSP 2024-2024 IEEE International Conference on Acoustics, Speech and Signal Processing (ICASSP)},
  pages={10761--10765},
  year={2024},
  organization={IEEE}
}

@article{khan2023battling,
  title={Battling voice spoofing: a review, comparative analysis, and generalizability evaluation of state-of-the-art voice spoofing counter measures},
  author={Khan, Awais and Malik, Khalid Mahmood and Ryan, James and Saravanan, Mikul},
  journal={Artificial Intelligence Review},
  volume={56},
  number={Suppl 1},
  pages={513--566},
  year={2023},
  publisher={Springer}
}

@misc{lowy_audio_deepfakes,
  author       = {Lowy Institute},
  title        = {Don't Play it by Ear: Audio Deepfakes and the Year of Global Elections},
  year         = {2023},
  url          = {https://www.lowyinstitute.org/the-interpreter/don-t-play-it-ear-audio-deepfakes-year-global-elections},
  note         = {Accessed: 2026-03-14}
}

@inproceedings{yi2022add,
  author    = {Jiangyan Yi and Ruibo Fu and Jianhua Tao and Shuai Nie and Haoxin Ma and Chenglong Wang and Tao Wang and Zhengkun Tian and Ye Bai and Cunhang Fan and others},
  title     = {{ADD} 2022: The First Audio Deep Synthesis Detection Challenge},
  booktitle = {Proc. ICASSP},
  pages     = {9216--9220},
  year      = {2022}
}

@article{yi2023add,
  author    = {Jiangyan Yi and Jianhua Tao and Ruibo Fu and Xiaohai Tian and Chenglong Wang and Tao Wang and Chu-Yuan Zhang and Xinrui Zhang and Yan Zhao and Yong Ren and others},
  title     = {{ADD} 2023: The Second Audio Deepfake Detection Challenge},
  journal   = {arXiv preprint arXiv:2305.13774},
  year      = {2023}
}

@inproceedings{baevski2020wav2vec,
  author    = {Alexei Baevski and Yuhao Zhou and Abdelrahman Mohamed and Michael Auli},
  title     = {wav2vec 2.0: A Framework for Self-Supervised Learning of Speech Representations},
  booktitle = {Advances in Neural Information Processing Systems (NeurIPS)},
  volume    = {33},
  pages     = {12449--12460},
  year      = {2020}
}

@article{chen2022wavlm,
  author    = {Sanyuan Chen and Chengyi Wang and Zhengyang Chen and Yu Wu and Shujie Liu and Zhuo Chen and Jinyu Li and Naoyuki Kanda and Takuya Yoshioka and Xiong Xiao and others},
  title     = {{WavLM}: Large-Scale Self-Supervised Pre-Training for Full Stack Speech Processing},
  journal   = {IEEE Journal of Selected Topics in Signal Processing},
  volume    = {16},
  number    = {6},
  pages     = {1505--1518},
  year      = {2022}
}

@article{hsu2021hubert,
  author    = {Wei-Ning Hsu and Benjamin Bolte and Yao-Hung Hubert Tsai and Kushal Lakhotia and Ruslan Salakhutdinov and Abdelrahman Mohamed},
  title     = {{HuBERT}: Self-Supervised Speech Representation Learning by Masked Prediction of Hidden Units},
  journal   = {IEEE/ACM Transactions on Audio, Speech, and Language Processing},
  volume    = {29},
  pages     = {3451--3460},
  year      = {2021}
}

@inproceedings{radford2023whisper,
  author    = {Alec Radford and Jong Wook Kim and Tao Xu and Greg Brockman and Christine McLeavey and Ilya Sutskever},
  title     = {Robust Speech Recognition via Large-Scale Weak Supervision},
  booktitle = {Proc. ICML},
  pages     = {28492--28518},
  year      = {2023}
}

@inproceedings{farooq2025transferable,
  title={Transferable adversarial attacks on audio deepfake detection},
  author={Farooq, Muhammad Umar and Khan, Awais and Uddin, Kutub and Malik, Khalid Mahmood},
  booktitle={Proceedings of the Winter Conference on Applications of Computer Vision},
  pages={1640--1649},
  year={2025}
}

@inproceedings{zhang2021lcnn,
  author    = {Lin Zhang and Xin Wang and Erica Cooper and Jose Patino and Nicholas Evans},
  title     = {An Initial Investigation for Detecting Partially Spoofed Audio},
  booktitle = {Proc. Interspeech},
  pages     = {4264--4268},
  year      = {2021}
}

@inproceedings{xie2024tdl,
  author    = {Yuankun Xie and Haonan Cheng and Yutian Wang and Long Ye},
  title     = {An Efficient Temporary Deepfake Location Approach Based Embeddings for Partially Spoofed Audio Detection},
  booktitle = {Proc. ICASSP},
  pages     = {966--970},
  year      = {2024}
}

@article{mao2026localizing,
  title={Localizing Speech Deepfakes Beyond Transitions via Segment-Aware Learning},
  author={Mao, Yuchen and Huang, Wen and Qian, Yanmin},
  journal={arXiv preprint arXiv:2601.21925},
  year={2026}
}

@article{zhong2024bam,
  author    = {Jie Zhong and Biao Li and Jiangyan Yi},
  title     = {Enhancing Partially Spoofed Audio Localization with Boundary-Aware Attention Mechanism},
  journal   = {arXiv preprint arXiv:2407.21611},
  year      = {2024}
}

@article{li2025tdam,
  author    = {Menglu Li and Lian Zhao and Xiao-Ping Zhang},
  title     = {Frame-Level Temporal Difference Learning for Partial Deepfake Speech Detection},
  journal   = {IEEE Signal Processing Letters},
  volume    = {32},
  pages     = {3053--3057},
  year      = {2025}
}

@inproceedings{anshul2026av,
  title={AV Representation Learning via Audio Shift Prediction for Multimodal Deepfake Detection and Temporal Localization},
  author={Anshul, Ashutosh and Chng, Eng Siong and Rajan, Deepu},
  booktitle={Proceedings of the IEEE/CVF Winter Conference on Applications of Computer Vision},
  pages={2553--2563},
  year={2026}
}

@inproceedings{luong2025llamapartialspoof,
  title={Llamapartialspoof: An llm-driven fake speech dataset simulating disinformation generation},
  author={Luong, Hieu-Thi and Li, Haoyang and Zhang, Lin and Lee, Kong Aik and Chng, Eng Siong},
  booktitle={ICASSP 2025-2025 IEEE International Conference on Acoustics, Speech and Signal Processing (ICASSP)},
  pages={1--5},
  year={2025},
  organization={IEEE}
}

@misc{Had-dataset,
      title={Half-Truth: A Partially Fake Audio Detection Dataset}, 
      author={Jiangyan Yi and Ye Bai and Jianhua Tao and Haoxin Ma and Zhengkun Tian and Chenglong Wang and Tao Wang and Ruibo Fu},
      year={2023},
      eprint={2104.03617},
      archivePrefix={arXiv},
      primaryClass={cs.SD},
      url={https://arxiv.org/abs/2104.03617}, 
}

@inproceedings{martin2023vicomtech,
  author    = {Juan Manuel Mart{\'i}n-Do{\~n}as and Aitor {\'A}lvarez},
  title     = {The {Vicomtech} Partial Deepfake Detection and Location System for the 2023 {ADD} Challenge},
  booktitle = {Proc. IJCAI Workshop on Deepfake Audio Detection and Analysis (DADA)},
  pages     = {37--41},
  year      = {2023}
}

@inproceedings{xian2025nepadd,
  author    = {Huhong Xian and Rui Liu and Berrak Sisman and Haizhou Li},
  title     = {{NE-PADD}: Leveraging Named Entity Knowledge for Robust Partial Audio Deepfake Detection via Attention Aggregation},
  booktitle = {Proc. APSIPA ASC},
  pages     = {2199--2204},
  year      = {2025}
}

@article{zhang2021initial,
  title={An initial investigation for detecting partially spoofed audio},
  author={Zhang, Lin and Wang, Xin and Cooper, Erica and Yamagishi, Junichi and Patino, Jose and Evans, Nicholas},
  journal={arXiv preprint arXiv:2104.02518},
  year={2021}
}

@article{zhang2021multi,
  title={Multi-task learning in utterance-level and segmental-level spoof detection},
  author={Zhang, Lin and Wang, Xin and Cooper, Erica and Yamagishi, Junichi},
  journal={arXiv preprint arXiv:2107.14132},
  year={2021}
}

@article{alali2025partial,
  author    = {Abdulazeez Alali and George Theodorakopoulos},
  title     = {Partial Fake Speech Attacks in the Real World Using Deepfake Audio},
  journal   = {Journal of Cybersecurity and Privacy},
  volume    = {5},
  number    = {6},
  year      = {2025}
}

@article{zhang2025partialedit,
  title={PartialEdit: identifying partial deepfakes in the era of neural speech editing},
  author={Zhang, You and Tian, Baotong and Zhang, Lin and Duan, Zhiyao},
  journal={arXiv preprint arXiv:2506.02958},
  year={2025}
}

@inproceedings{ge2025gncl,
  title={GNCL: A graph neural network with consistency loss for segment-level spoofed speech detection},
  author={Ge, Zirui and Xu, Xinzhou and Guo, Haiyan and Yang, Zhen and Schuller, Bj{\"o}rn},
  booktitle={ICASSP 2025-2025 IEEE International Conference on Acoustics, Speech and Signal Processing (ICASSP)},
  pages={1--5},
  year={2025},
  organization={IEEE}
}

@article{li2025survey,
  title={A survey on speech deepfake detection},
  author={Li, Menglu and Ahmadiadli, Yasaman and Zhang, Xiao-Ping},
  journal={ACM Computing Surveys},
  volume={57},
  number={7},
  pages={1--38},
  year={2025},
  publisher={ACM New York, NY}
}

@article{he2025manipulated,
  title={Manipulated Regions Localization For Partially Deepfake Audio: A Survey},
  author={He, Jiayi and Yi, Jiangyan and Tao, Jianhua and Zeng, Siding and Gu, Hao},
  journal={arXiv preprint arXiv:2506.14396},
  year={2025}
}

@article{fu2025multi,
  title={Multi-Layer {WavLM} Based Partial Deepfake Speech Detection Using Frame-Level Feature Engineering},
  author={Fu, Jingchang and Yeung, Siu-Kei Au and Hung, Kevin},
  journal={arXiv preprint},
  year={2025}
}

@inproceedings{dragar2024w,
  title={{W-TDL}: Window-Based Temporal Deepfake Localization},
  author={Dragar, Luka and Rot, Peter and Peer, Peter and {\v{S}}truc, Vitomir and Batagelj, Borut},
  booktitle={Proceedings of the 2nd International Workshop on Multimodal and Responsible Affective Computing},
  pages={24--29},
  year={2024}
}

@article{yang2026forensic,
  title={Forensic deepfake audio detection using segmental speech features},
  author={Yang, Tianle and Sun, Chengzhe and Lyu, Siwei and Rose, Phil},
  journal={Forensic Science International},
  pages={112768},
  year={2026},
  publisher={Elsevier}
}

@article{cao2025sail,
  author    = {Jialu Cao and Hui Tian and Peng Tian and Haizhou Li and Jianzong Wang},
  title     = {Robust Detection of Partially Spoofed Audio Using Semantic-Aware Inconsistency Learning},
  journal   = {IEEE/ACM Transactions on Audio, Speech, and Language Processing},
  year      = {2025}
}

@article{liu2025zeroday,
  author    = {Xuechen Liu and Xin Wang and Junichi Yamagishi},
  title     = {Zero-Day Audio DeepFake Detection via Retrieval Augmentation and Profile Matching},
  journal   = {arXiv preprint arXiv:2509.21728},
  year      = {2025}
}

@inproceedings{stan2025tada,
  author    = {Robert Stan and others},
  title     = {{TADA}: Training-free Attribution and Out-of-Domain Detection of Audio Deepfakes},
  booktitle = {Proc. Interspeech},
  year      = {2025}
}

@article{liu2025nes2net,
  author    = {Tianchi Liu and Duc-Tuan Truong and Rohan Kumar Das and Kong Aik Lee and Haizhou Li},
  title     = {{Nes2Net}: A Lightweight Nested Architecture for Foundation Model Driven Speech Anti-Spoofing},
  journal   = {IEEE Transactions on Information Forensics and Security},
  year      = {2025},
  publisher = {IEEE}
}

@inproceedings{lv2022fake,
  title={Fake audio detection based on unsupervised pretraining models},
  author={Lv, Zhiqiang and Zhang, Shanshan and Tang, Kai and Hu, Pengfei},
  booktitle={ICASSP 2022-2022 IEEE International Conference on Acoustics, Speech and Signal Processing (ICASSP)},
  pages={9231--9235},
  year={2022},
  organization={IEEE}
}

@inproceedings{tak2021end,
  title={End-to-end anti-spoofing with {RawNet2}},
  author={Tak, Hemlata and Patino, Jose and Todisco, Massimiliano and Nautsch, Andreas and Evans, Nicholas and Larcher, Anthony},
  booktitle={ICASSP 2021-2021 IEEE International Conference on Acoustics, Speech and Signal Processing (ICASSP)},
  pages={6369--6373},
  year={2021},
  organization={IEEE}
}

@article{li2025audio,
  title={Audio Deepfake Detection via a Fuzzy Dual-Path Time-Frequency Attention Network},
  author={Li, Jinzi and Wang, Hexu and Xie, Fei and Feng, Xiaozhou and Chen, Jiayao and Liu, Jindong and Wang, Juan},
  journal={Sensors},
  volume={25},
  number={24},
  pages={7608},
  year={2025},
  publisher={MDPI}
}

@article{obi2023review,
  title={A review of techniques for regularization},
  author={Obi, Jude Chukwura and Jecinta, Ibebuike Chinenye},
  journal={International Journal of Research in Engineering and Science},
  volume={11},
  number={1},
  pages={360--367},
  year={2023}
}

@article{van2017l2,
  title={L2 regularization versus batch and weight normalization},
  author={Van Laarhoven, Twan},
  journal={arXiv preprint arXiv:1706.05350},
  year={2017}
}

@inproceedings{uddin2025sheild,
  title={Sheild: A secure and highly enhanced integrated learning for robust deepfake detection against adversarial attacks},
  author={Uddin, Kutub and Khan, Awais and Farooq, Muhammad Umar and Malik, Khalid Mahmood},
  booktitle={Proceedings of the IEEE/CVF International Conference on Computer Vision},
  pages={1502--1511},
  year={2025}
}

@article{uddin2025advbench,
  title={AdvBench: A Comprehensive Benchmark of Adversarial Attacks on Deepfake Detectors in Real-World Consumer Applications},
  author={Uddin, Kutub and Farooq, Muhammad Umar and Khan, Awais and Saeed, Muhammad Saad and Haq, Ijaz Ul and Tasnim, Nusrat and Malik, Khalid Mahmood},
  journal={Authorea Preprints},
  year={2025},
  publisher={Authorea}
}

@article{uddin2025adversarial,
  title={Adversarial attacks on audio deepfake detection: A benchmark and comparative study},
  author={Uddin, Kutub and Farooq, Muhammad Umar and Khan, Awais and Malik, Khalid Mahmood},
  journal={arXiv preprint arXiv:2509.07132},
  year={2025}
}
}
% \input{sec/supplementary_Matrials}
\end{document}